\documentclass[aps,floats,twocolumn,superscriptaddress]{revtex4}

\usepackage{amsmath,bm,graphicx}

\def\P{{\mathcal P}}

\def\E{{\mathcal E}}
\def\N{{\mathcal N}}

\begin{document}

\title{Microwave power generated by a spin-torque oscillator in the presence of noise}

\author{Vasil Tiberkevich}
\affiliation{Department of Physics, Oakland University, Rochester, MI 48309, USA}
\author{Joo-Von Kim}
\affiliation{Institut d'Electronique Fondamentale, UMR CNRS 8622, Universit{\'e} Paris-Sud, 91405 Orsay cedex, France}
\author{Andrei Slavin}
\affiliation{Department of Physics, Oakland University, Rochester, MI 48309, USA}

\date{\today}

\begin{abstract}
An expression for the microwave power generated by a spin-torque
oscillator in the presence of thermal noise is derived. This
expression, when used in a subcritical ($I<I_{\rm th}$) regime,
demonstrates that generated power is determined by the noise level
and provides a simple way to experimentally determine the threshold
current $I_{\rm th}$ for microwave generation. The same expression gives
a good quantitative description of the experimentally measured
dependence of the generated power on the bias current $p(I)$ in a
moderate super-critical ($I_{\rm th} < I < 1.5 I_{\rm th}$) range of  bias
current variation.
\end{abstract}

\maketitle
It has been demonstrated both theoretically
\cite{slonczewski96, berger96} and experimentally
\cite{tsoi98,kiselev03, rippard04a} that the spin-torque effect in
magnetic nanostructures creates an effective negative magnetic
damping, which for sufficiently large bias currents ($I>I_{\rm th}$) can
compensate positive damping. This leads to the excitation of one of the spin wave
eigenmodes of the structure, thus providing a way to create fully
metallic nano-sized  microwave spin-torque  oscillators (STO).

In contrast with almost all other conventional microwave oscillators, STOs are very small in size and, therefore, the influence of thermal fluctuations (thermal noise) on their operation is of a critical
importance. This problem bears some similarity with the case of microwave oscillators based of
Josephson junctions \cite{likharev86}. An STO in the sub-critical regime ($I<I_{\rm th}$) in the presence of thermal noise, in particular, functions as a regenerative filter. With the increase of bias current in this regime (which corresponds to a decrease of net magnetic damping), the oscillator filters effectively its lowest eigenfrequency from noise, thus providing a measurable output microwave signal at
this frequency much before the threshold of self-sustained microwave generation ($I=I_{\rm th}$) is attained. As such, an experimental determination of the STO threshold current in the presence of thermal noise presents a challenge, which can only be met if the influence of thermal noise on the STO operation is
accounted for in theory.

In this Letter we present a theory of microwave generation in STOs in which the influence of thermal noise is explicitly taken into account, in contrast with previously developed theories~\cite{rezende05a,slavin05a}. The developed spin-wave theory of microwave generation is valid for a wide range of bias currents: from zero current at which the system is in the state of thermal equilibrium, to a significantly above-threshold (super-critical) currents at which the influence of thermal noise can be neglected. The theory gives good quantitative description of recent experimental results on microwave power generated by a current-driven magnetic nanopillar \cite{mistral06a}, and also allows for a simple and accurate procedure for determining the threshold current of self-sustained microwave generation in STOs from the measured dependence of the output microwave power on the bias current in the {\em subcritical} regime.

It has been demonstrated in Refs.~\onlinecite{rezende05a,slavin05a} that the magnetization dynamics in STOs
can be adequately described by the dimensionless complex amplitude $c(t)$ of a \textit{single} spin wave mode excited by the bias direct current. The phase $\phi \equiv \arg(c)$ of the complex amplitude $c(t)$ is equal to the azimuthal angle of the magnetization precession, while the oscillation power $p \equiv |c|^2$ determines the polar angle of the precession, $\theta \equiv \arccos(M_z/M_0) = \arccos(1 - 2p)$. Here $M_0$ is the length of the magnetization vector in the ``free" magnetic layer, and $M_z$ is the
projection of this vector on the direction of stationary equilibrium magnetization $\bf z$ (see Ref.~\onlinecite{slavin05a} for details).

The equation of motion for the amplitude $c(t)$ in the general case, whereby the thermal noise is taken into account, can be written as
\begin{equation}\label{model}
    \frac{\partial c}{\partial t} + i\omega(p)c + \Gamma_+(p)c - \Gamma_-(p)c = f(t)
\,,\end{equation} 
where $\omega(p)$ is the frequency of the excited spin wave mode, $\Gamma_+(p)$ is the natural positive damping in magnetic system, $\Gamma_-(p)$ is the effective negative damping introduced into the system by the spin-polarized current, and $f(t)$ is a random white Gaussian process that describes the influence of the thermal fluctuations (noise). The correlation function of this random noise $f(t)$ can be written as
\begin{equation}
    \langle f^*(t)f(t') \rangle = 2 D \delta(t-t')
\ ,\end{equation} 
where $D$ is the diffusion coefficient that characterizes the noise amplitude.

In general, the frequency $\omega(p)$ and the effective damping rates $\Gamma_\pm(p)$ are complicated nonlinear functions of the spin wave power $p$. For moderate levels of excitations of the spin system, however, one can use a Taylor series expansion of these functions and write approximately (see, e.g., Refs. \onlinecite{phaselock06} and \onlinecite{NonlinearGilbert}):
\begin{subequations}\label{pars}
\begin{eqnarray}\label{pars-a}
    \omega(p) &\approx& \omega_0 + N p
\,,\\\label{pars-b}
    \Gamma_+(p) \approx \Gamma_0(1 + Q p)
\,,\ &&\ 
    \Gamma_-(p) \approx \sigma I(1 - p)
\ .\end{eqnarray}
\end{subequations}
Here $\omega_0$ is the linear oscillation frequency and $N$ is the nonlinear frequency shift coefficient (see Eq.~(38) in Ref.~\onlinecite{slavin05a}), $\Gamma_0$ is the linear damping of the excited spin wave mode in the passive regime, $Q$ is a phenomenological coefficient characterizing the nonlinearity of positive damping (see Ref.~\onlinecite{NonlinearGilbert} for details),  $\sigma$ is the spin-polarization efficiency defined in Eq.~(2) of Ref.~\onlinecite{phaselock06}, and $I$ is the bias charge current.

We would like to stress that Eq.~(\ref{model}) can adequately describe an auto-oscillator of {\em any} nature under the influence of white noise $f(t)$, provided that this oscillator has a nonlinear frequency $\omega(p)$, nonlinear natural positive damping $\Gamma_+(p)$, and nonlinear negative damping $\Gamma_-(p)$, where $p = |c|^2$ is the oscillation power.

It should also be noted that the complex spin wave amplitude $c$ is a {\em canonical} complex variable of the magnetic system. In particular, this means that the phase volume $d\Omega$ of some element of the system's phase space is simply proportional to the area of this element on the complex $c$-plane, i.e., in power-phase coordinates, $d\Omega = dp\, d\phi$. As such, the probability distribution function $\P_{\rm eq}(p, \phi)$ of the system at thermal equilibrium (i.e. for $\Gamma_-(p) = 0$) should have the form of the pure Boltzmann exponential distribution,
\begin{equation}\label{P-eq}
    \P_{\rm eq}(p, \phi) \propto \exp\left(-\frac{\E(p)}{k_{\rm B}T}\right)
\,,\end{equation} without any pre-exponential factors. In
Eq.~(\ref{P-eq}) $k_{\rm B}$ is the Boltzmann constant, $T$ is the
absolute temperature, and $\E(p)$ is the energy of the system,
\begin{equation}\label{energy}
    \E(p) = \lambda \int_0^p \omega(p') dp' \approx \lambda \left(\omega_0 p +\frac12 N p^2\right)
\ .\end{equation}
Here the constant $\lambda$ depends on the normalization of the amplitude $c$ and, in the case of an STO, is given by 
\begin{equation}\label{lambda}
    \lambda = V_{\rm eff}M_0/\gamma
\,,\end{equation}
where $V_{\rm eff}$ is the effective volume of the magnetic material of the ``free" layer involved in the auto-oscillation, and $\gamma$ is the gyromagnetic ratio.

To describe correctly the stochastic dynamics of a \textit{nonlinear} oscillator with arbitrary dependences of the frequency $\omega = \omega(p)$ and natural damping $\Gamma_+ =
\Gamma_+(p)$ on the oscillation power $p$, one has to assume that the diffusion coefficient $D$ also depends on $p$ as
\begin{equation}\label{Dn}
    D(p) = \Gamma_+(p)\eta(p)= \Gamma_+(p)\frac{k_{\rm B}T}{\lambda\omega(p)}
\ ,\end{equation} 
where $\eta(p)$ is the effective noise power in the nonlinear regime. As it will be shown below, this form of the diffusion coefficient provides correct equilibrium distribution Eq.~(\ref{P-eq}) for {\em arbitrary} power dependences $\omega(p)$ and $\Gamma_+(p)$.

The stochastic Langevin Eq.~(\ref{model}) produces the following deterministic Fokker-Planck equation for the probability distribution function $\P(t, p, \phi)$ of the system,
\begin{eqnarray}\label{FP}
    \frac{\partial\P}{\partial t}  &=&
            \frac{\partial}{\partial p}\left[2p\left(\Gamma_+ - \Gamma_-\right)\P\right]
            +\omega\frac{\partial\P}{\partial\phi}
\\\nonumber&&
            +\frac{\partial}{\partial p}\left(2pD\frac{\partial\P}{\partial p}\right)
            +\frac{\partial}{\partial\phi}\left(\frac{D}{2p}\frac{\partial\P}{\partial\phi}\right)
\ .\end{eqnarray}

While the general non-stationary solution of the Fokker-Planck Eq.~(\ref{FP}) is rather complicated, it is possible to find a simple analytic expression for the \textit{stationary} probability distribution function $\P_0(p)$ that is independent of the oscillator phase $\phi$, since in the stationary state all the phases of the oscillator are equally possible. This expression has the form,
\begin{equation}\label{P0}
    \P_0(p) = \N_0 \exp\left[
            -\frac{\lambda}{k_{\rm B}T} \int_0^p \omega(p')\left(1 - \frac{\Gamma_-(p')}{\Gamma_+(p')}\right)dp'
        \right]
\,,\end{equation} where $\N_0$ is the normalization constant determined from the normalization condition $ \int_0^\infty \P_0(p) dp = 1 $.

From Eq.~(\ref{P0}) one can see by inspection that our approach gives the standard Boltzmann distribution Eq.~(\ref{P-eq}) at thermal equilibrium ($\Gamma_-(p) = 0$). Eq.~(\ref{P0}) allows one to determine any stationary characteristic of the STO.  For instance, the mean oscillation power $\overline p$ is found to be
\begin{equation}\label{pmean}
\overline p = \int_0^\infty p\, \P_0(p) dp
\ .\end{equation}

\begin{figure}
\includegraphics[width=0.4\textwidth]{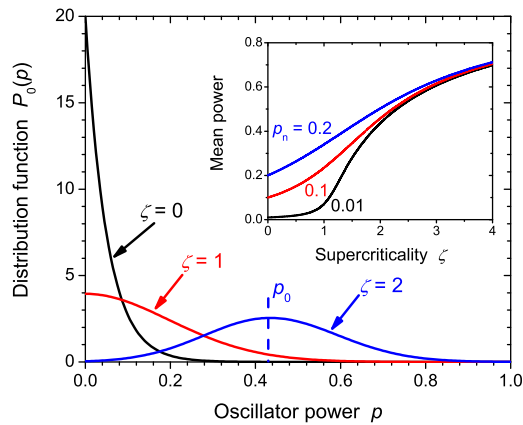}
\caption{\label{f:PDF} Main panel: Probability distribution function $\P_0(p)$ (see Eq.~(\ref{P0})) for several supercriticalities $\zeta$. Nonlinear damping coefficient $Q = 0.3$, noise level $\eta = 0.05$. Inset: Dependence of the mean oscillator power $\overline p$ (see Eq.~(\ref{pmean})) on the supercriticality $\zeta$ for several noise levels $\eta$. Nonlinear damping coefficient $Q = 0.3$. }
\end{figure}

Equations (\ref{P0}) and (\ref{pmean}) allow one to calculate the probability distribution function $\P_0(p)$  and the mean microwave power $\overline p$ generated by a STO  in the presence of thermal noise and nonlinearity of the damping. The results of calculations performed using these equations are presented in Fig.~\ref{f:PDF}. The main panel of this figure illustrates the probability distribution function Eq.~(\ref{P0}) for several values of the supercriticality parameter $\zeta \equiv I/I_{\rm th}$, where $I_{\rm th} \equiv \Gamma_0/\sigma$ is the critical current at which self-sustained oscillations start.

In the sub-critical regime ($\zeta < 1$) the distribution $\P_0(p)$ has a maximum at $p = 0$. Expanding the integrand in Eq.~(\ref{P0}) in Taylor series near $p = 0$, and keeping only the first non-zero term, one can obtain approximate expressions for the distribution $\P_0(p) \sim \exp\left[-(1 - \zeta)p/\eta(0)\right]$ and mean power in this regime,
\begin{equation}\label{pmean-sub}
    \overline p \simeq \frac{\eta(0)}{1 - \zeta}
            = \left(\frac{k_{\rm B}T}{\lambda\omega_0}\right)\frac{I_{\rm th}}{I_{\rm th} - I}
\ .\end{equation}

In the super-critical regime ($\zeta > 1$), the effective negative damping $\Gamma_-(p)$ compensates the equilibrium positive damping $\Gamma_+(p)$ at a certain oscillator
power $p_0$. At this power, $\Gamma_-(p_0) = \Gamma_+(p_0)$, and from
the approximate Eq.~(\ref{pars}) one obtains,
\begin{equation}\label{pmean-super}
     p_0 = \frac{\zeta - 1}{\zeta + Q}=\frac{I-I_{\rm th}}{I+QI_{\rm th}}
\ .\end{equation} 
The distribution function $\P_0$ has a maximum at $p = p_0$ (see Eq.~(\ref{P0})). Thus, by expanding the integrand in Eq.~(\ref{P0}) in a Taylor series near $p = p_0$, one can derive an approximate expression for $\P_0(p) \sim \exp\left[-(p-p_0)^2/2\Delta p^2\right]$ which is valid in the super-critical regime. Here $\Delta p^2 = \eta(p_0)(1+Q)\zeta/(\zeta+Q)^2$ is the
level of power fluctuations of STO in the super-critical regime. The mean oscillation power $\overline p$ in this regime is approximately equal to $p_0$ (see Eq.~(\ref{pmean-super})).

The main panel in Fig.~\ref{f:PDF} illustrates the probability distribution function Eq.(\ref{P0}) in the subcritical $(\zeta=0)$, critical $(\zeta=1)$, and super-critical ($\zeta=2$) regimes. The inset in Fig.~\ref{f:PDF} shows  that the average power generated by
STO (determined by Eq.~(\ref{pmean})) is substantially influenced by
the power of thermal noise.

\begin{figure}
\includegraphics[width=0.4\textwidth]{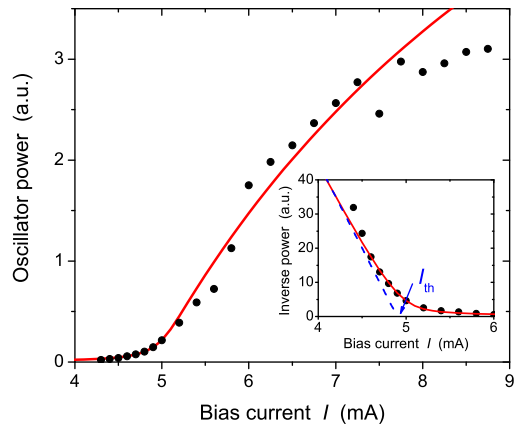}
\caption{\label{f:power} Main panel: Dependence of the mean power $\overline p$ on the bias current $I$. Dots -- experiment of Ref.~\onlinecite{mistral06a} for $T = 225$~K, solid line -- theoretical
dependence (\ref{pmean}) for $I_{\rm th} = 4.9$~mA, $\eta = 4.2\cdot10^{-4}$, $Q = 0.3$. Inset shows the same data for {\em inverse} power $1/\overline p$ in near-threshold range of currents. Dashed line corresponds to the approximate expression Eq.~(\ref{pmean-sub}) valid for small currents. Intersection of this line this $x$-axis gives the value of the threshold current $I_{\rm th}$. }
\end{figure}

As an illustration of our approach we compare the obtained theoretical results with recent measurements of the generated power in current-driven metallic nanopillars \cite{mistral06a}. In the inset of Fig.~\ref{f:power} we show the dependence of the {\em inverse} mean power $1/\overline p$ on the bias current $I$. According to Eq.~(\ref{pmean-sub}), this dependence is linear for small values of the bias current, $1/\overline p \propto (I_{\rm th} - I)$. This linear dependence in the subcritical regime $I < I_{\rm th}$ allows a method of determining precisely the threshold current $I_{\rm th}$ for microwave generation in STO for situations in which the influence of the thermal fluctuations is strong enough and determination of $I_{\rm
th}$ by other means is difficult. Thus, from the inset of Fig.~\ref{f:power} one can immediately deduce the value of the threshold current $I_{\rm th} = 4.9$~mA, which is substantially larger than the current $I_* \approx 4.2$~mA, at which the thermally-induced oscillations become observable in the experiment \cite{mistral06a}.

Using the value of the threshold current $I_{\rm th} = 4.9$~mA, derived from the experimental data in the deeply subcritical regime and Eq(\ref{pmean}), it is possible to describe the  power generated
by STO $\overline p$ in a wide range of bias currents (see main panel in Fig.~\ref{f:power}).

In conclusion, we have developed a simple theory of microwave generation in STOs based on a stochastic nonlinear oscillator model. We demonstrate that this theory allows one to accurately determine the threshold current for microwave generation from experimental data and quantitatively describes  the experimentally measured STO power for moderate ($I\leq1.5 I_{\rm th}$) values of the bias current.

This work was in part supported by the MURI grant W911NF-04-1-0247 from the US Army Research Office, by the contract No.W56HZV-07-P-L612 from the U.S. Army TARDEC, RDECOM, and by the
grant ECCS-0653901 from the National Science Foundation. JK acknowledges support from the European Communities program IST under the Contract No. IST-016939 TUNAMOS.

\clearpage

\end{document}